 \documentclass[sn-mathphys,Numbered]{sn-jnl}% Math and Physical Sciences Reference Style
\usepackage{graphicx}%
\usepackage{multirow}%
\usepackage{amsmath,amssymb,amsfonts}%
\usepackage{amsthm}%
\usepackage{mathrsfs}%
\usepackage[title]{appendix}%
\usepackage{xcolor}%
\usepackage{textcomp}%
\usepackage{manyfoot}%
\usepackage{booktabs}%
\usepackage{algorithm}%
\usepackage{algorithmicx}%
\usepackage{algpseudocode}%
\usepackage{listings}%
\begin{document}

\title[Article Title]{Hyper Spectral Resolution Stimulated Raman Spectroscopy with Amplified fs Pulse Bursts}

%%=============================================================%%
%% Prefix	-> \pfx{Dr}
%% GivenName	-> \fnm{Joergen W.}
%% Particle	-> \spfx{van der} -> surname prefix
%% FamilyName	-> \sur{Ploeg}
%% Suffix	-> \sfx{IV}
%% NatureName	-> \tanm{Poet Laureate} -> Title after name
%% Degrees	-> \dgr{MSc, PhD}
%% \author*[1,2]{\pfx{Dr} \fnm{Joergen W.} \spfx{van der} \sur{Ploeg} \sfx{IV} \tanm{Poet Laureate} 
%%                 \dgr{MSc, PhD}}\email{iauthor@gmail.com}
%%=============================================================%%

\author[1]{\fnm{Hongtao} \sur{Hu}}

\author[1]{\fnm{Tobias} \sur{Flöry}}

\author[1]{\fnm{Vinzenz} \sur{Stummer}}

\author[1]{\fnm{Audrius} \sur{Pugzlys}}

\author[1]{\fnm{Markus} \sur{Kitzler-Zeiler}}

\author*[2]{\fnm{Xinhua} \sur{Xie}}\email{xinhua.xie@psi.ch}

\author[3]{\fnm{Alexei} \sur{Zheltikov}}

\author*[1]{\fnm{Andrius} \sur{Baltu\v{s}ka}}\email{andrius.baltuska@tuwien.ac.at}

\affil[1]{\orgdiv{Photonics Institute}, \orgname{Technische Universität Wien}, \orgaddress{\street{Gußhausstraße 27-29}, \city{Vienna}, \postcode{A-1040}, \country{Austria}}}

\affil[2]{\orgdiv{SwissFEL}, \orgname{Paul Scherrer Institute,} \orgaddress{\city{Villigen PSI}, \postcode{5232},  \country{Switzerland}}}

\affil[3]{\orgdiv{Department of Physics and Astronomy}, \orgname{Texas A$\&$M University}, \orgaddress{\city{College Station}, \postcode{77843}, \state{Texas}, \country{USA}}}

%%==================================%%
%% sample for unstructured abstract %%
%%==================================%%

\abstract{We present a novel approach to achieve hyper spectral resolution, high sensitive detection, and high speed data acquisition Stimulated Raman Spectroscopy by employing amplified offset-phase controlled fs-pulse bursts.
In this approach, the Raman-shift spectrum is obtained through the direct mapping between the bursts offset phase and the Raman-shift frequency, which requires neither wavelength-detuning as in the long-pulse method nor precise dispersion management and delay scanning with movable parts as in the spectral focusing technique.
This method is demonstrated numerically by solving the coupled non-linear Schrödinger equations and the properties of this approach are systematically investigated.
The product of the spectral resolution and the pixel dwell time in this work is below $2 ~ \mu s\cdot cm^{-1}$, which is at least an order of magnitude lower than previous methods.
This previously untouched area will greatly expand the applications of SRS and holds the potential for discovering new science.}

%%================================%%
%% Sample for structured abstract %%
%%================================%%

\keywords{Stimulated Raman Spectroscopy, Femtosecond Pulse Bursts, Nonlinear Optics}

%%\pacs[JEL Classification]{D8, H51}

%%\pacs[MSC Classification]{35A01, 65L10, 65L12, 65L20, 65L70}

\maketitle

\section{Introduction}\label{sec1}

Stimulated Raman scattering (SRS) plays a prominent role in the resonant nonlinear spectroscopy by addressing Raman-active (and thus optically-inactive) vibrational and rotational transitions, distinguishing itself from spontaneous Raman scattering \cite{1928Raman} by providing 3$\sim$4 orders of magnitude stronger signals. 
Since the first observation \cite{1962Eckhardt,1965Maker}, coherent Raman scattering has been widely applied in biological imaging \cite{2015ReviewXie,2021ReviewLiu,2019ReviewValev}, environmental gas sensing \cite{2015ReviewXu,2011Hemmer,2015Malevich}, materials characterization \cite{Ferrari2013,Plechinger2012}, and other fields \cite{2020Mackonis,2005Rong,1976Hill,2018Polli,2013Ideguchi,2012Schliesser}.

 Scanning the Raman-shift frequency, the molecular vibrational or rotational energy levels, is the most essential and crucial part of Stimulated Raman scattering applications.
The choice of the driven laser fields to excite the molecules or materials is of vital importance since it fundamentally determines the spectral resolution, acquisition speed, and detection sensitivity.
Depending on whether picosecond (ps) or femtosecond (fs) laser pulses are used, current methods of SRS can be divided into three categories\cite{2021ReviewLiu}: sequential wavelength tuning (ps+ps) \cite{2008Xie,2012Suhalim,2013Kong,1982Duncan}, spectral focusing (fs+fs) \cite{2013Xie,2004Hellerer,2019Jolly}, and parallel multiplex excitation (ps+fs) \cite{2012Fu,2015Liao}.
Each method has its own advantage, e.g. the sequential wavelength tuning approach achieves a high spectral resolution down to $ 5 cm^{-1}$ \cite{2012Suhalim}; the spectral focusing method provides a high sensitivity of detection \cite{2004Hellerer}; the techniques based on parallel multiplex excitation have high data-collection speed up to $60~{\mu}s$ per frame \cite{2015Liao}. 
The highest data acquisition speed is $0.5 \mu s/\text{pixel}$ achieved by the dual-phase SRS method. \cite{2017He}.

In general, the advantages of the three categories SRS methods can not be achieved at the same time. For example, the conventional Raman shift scanning method in spectral focusing with changing the time delay of two linearly chirped fs pulses has a lower spectral resolution as compared to the method based on sequential wavelength tuning. On the other hand, the sequential wavelength detuning with high spectral resolution has a much lower data-acquisition speed than the parallel multiplex excitation method. 
In this work, we demonstrate a novel approach with burst-driven SRS (BSRS) to achieve those key advantages, high spectral resolution, high speed of data collection, and high sensitivity of detection, at the same time.

 BSRS using the amplified fs pulse bursts \cite{2020Vinzenz} to drive the SRS process can achieve a high spectral resolution of 0.2 $cm^{-1}$ and the short pixel dwell time of 10 $\mu s$. 
Moreover, the Raman shift scanning in this method requires neither wavelength detuning nor modulation frequency setting, nor precise dispersion management or a delay positioning system. 
It is accomplished by simply and digitally changing the offset phase of the pulse bursts. The offset phase refers to the phase difference between two neighboring pulses in the burst illustrated in Fig. \ref{fig:1} (a). 
Fig. \ref{fig:1} summarizes the working principle of Raman shift scanning.  
Specifically, the narrow bandwidth of each spike in the spectral domain is determined by the pulse number ($N$) and the inter-pulse temporal separation ($\Delta \tau$) between the adjacent pulses in the burst. 
And the frequency tuning is achieved through moving the spectral comb in the opposite direction when changing the offset phase of the burst [Fig. \ref{fig:1} (b)].
    % We also reveal the intrinsic dynamics of resonant and non-resonant Raman processes driven by bursts by analyzing the intensity difference of bursts in the time domain.
Combining a number of key advantages, this technology holds great promise for greatly improving and extending the applications of SRS.

\section{Results}\label{sec:scanning}

\subsection{Working principle}

      \begin{figure}[htbp]
    % \begin{figure*}[htbp]
    \centering
    
    \includegraphics[width=1\textwidth]{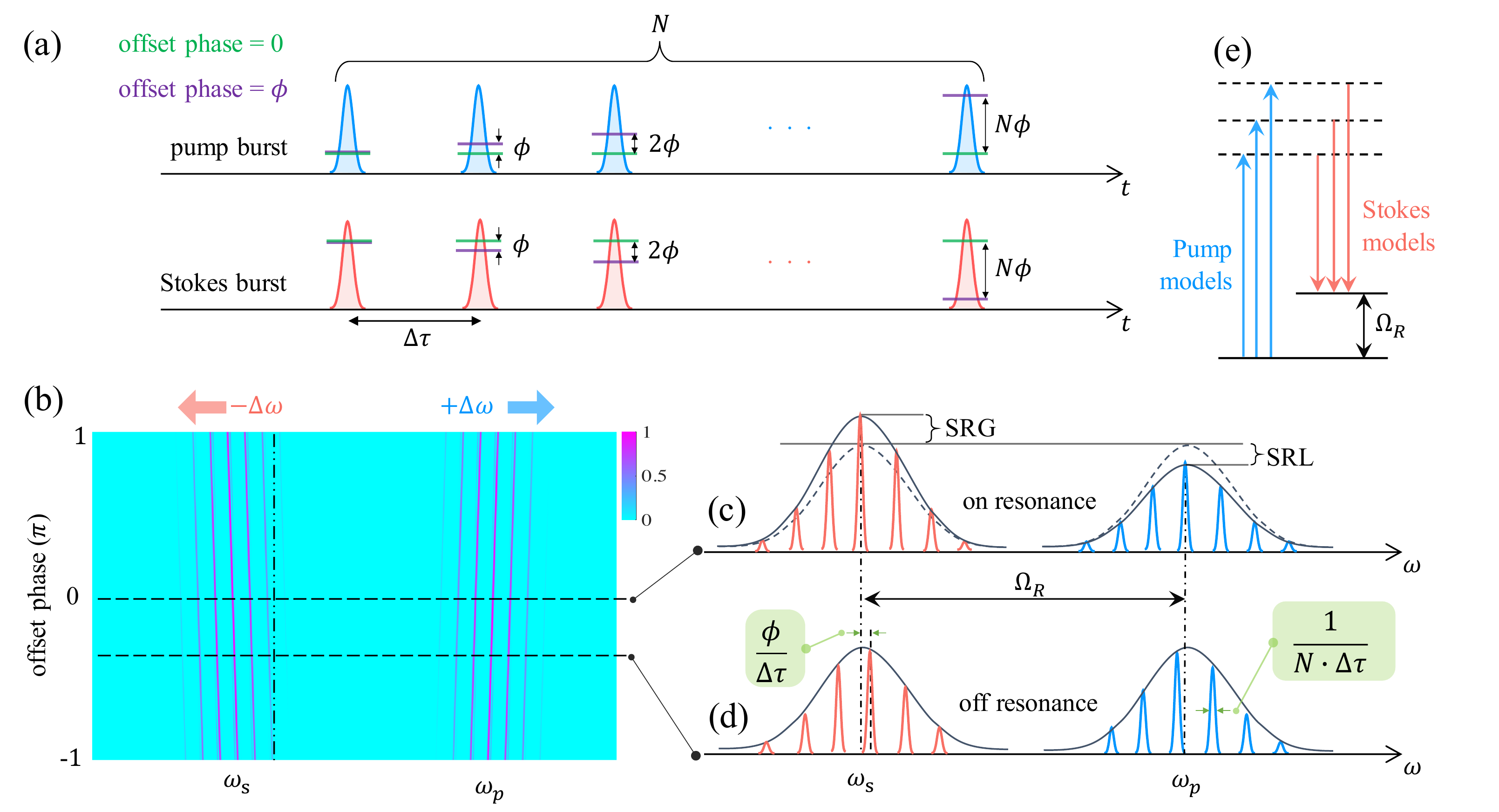}
    \caption{Working principle for Stimulated Raman Scattering driven by bursts. (a) Schematic diagram of the pump burst and the Stokes burst in the time domain with the offset phase equal to zero and $\phi$. (b) The comb peaks belonging to the pump burst and the Stokes burst move in the opposite direction due to phase conjugation when the offset phase changes. (c)  In the case of resonance, the comb peak frequency difference matches the Raman frequency, $\Omega_R$, leading to SRG (Stimulated Raman Gain) and SRL (Stimulated Raman Loss). The energy can flow from the pump burst to the Stokes burst via the vibrational excitation of a molecule.  (d) For the off-resonance case, the Raman process is strongly suppressed and leads to minor changes in the burst pulse energies depending on the detune. (e) Simultaneous stimulated Raman on-resonance is achieved by multiple pumps and Stokes comb frequencies.}
    \label{fig:1}
    % \end{figure*}
    \end{figure}

    The main technical prerequisite for implementing the method in Fig. \ref{fig:1} has been recently realized in our group \cite{2020Vinzenz} based on a programmable generation of amplified fs pulse bursts in a regenerative amplifier (RA) operated in the Vernier mode relative to the cavity length of the master oscillator (MO).
    As experimentally shown in Ref.  \cite{2020Vinzenz}, spectral interference of the pulses in the generated bursts leads to a comb-like structure with a 1 THz $\simeq$ 33.3 $cm^{-1}$ intermodal distance if the detuning between the RA and MO roundtrips $\Delta \tau = |L_{RA}-L_{MO}|/c = 1 ps$. 
    The THz-spaced pseudo-modes have the spectral width determined by the inverse duration of the burst, $(\Delta \tau\cdot N)^{-1}$, are filled with a very dense, MHz-interval-frequency comb structure \cite{Picque2019} that is irrelevant to our study. 
    The peak frequencies of the pseudo-modes, corresponding to constructive spectral interference, are programmed, alongside the individual amplitudes of the pulses forming the burst, by use of an acoustic-optic modulator (AOM) placed between the MO and RA.

     We note that the aspect of chirp management becomes irrelevant for the spectral narrowing achieved in this case purely through interference.
    Correspondingly, a white-light-seeded OPA driven by such a pulse burst will behave exactly like an OPA driven by a single isolated fs laser. 
    As a result, an independent signal-idler pulse pair will originate from each laser pulse in the burst and similar pseudo-combs will arise under the respective spectral envelopes (Fig. \ref{fig:1}) through the spectral interference of signal and idler pulse bursts. 
    We further note that due to phase conjugation in a properly designed OPA \cite{2002Baltuska}, the phase control of the fundamental laser burst will cause spectral modes to shift in the opposite directions for the signal and idler pseudo-combs. 
    Therefore, as shown in Fig. \ref{fig:1} (b) and (c), resonant SRS conditions can be simultaneously fulfilled or missed by multiple pseudo-modes depending on the input phase of the pulses loaded into the RA. 
    The free-space frequency scanning range corresponds to full the intermodal spacing, i.e. to ~33 $cm^{-1}$ under the assumed conditions, which is adequate to resolve a complete vibrational manifold structure of simple molecules (e.g. $N_2$) when the signal pulse burst is taken as an SRS pump wave and the idler burst – as a Stokes wave, respectively. A periodic jump to the next set of resonant pump-Stokes mode pairs is expected when the offset phase reaches an integer multiple of $\pi$.
    In the following analysis, we focus on the results of $N_2$, as an example.

    \begin{figure*}[ht!]
    \centering
    \includegraphics[width=1\textwidth]{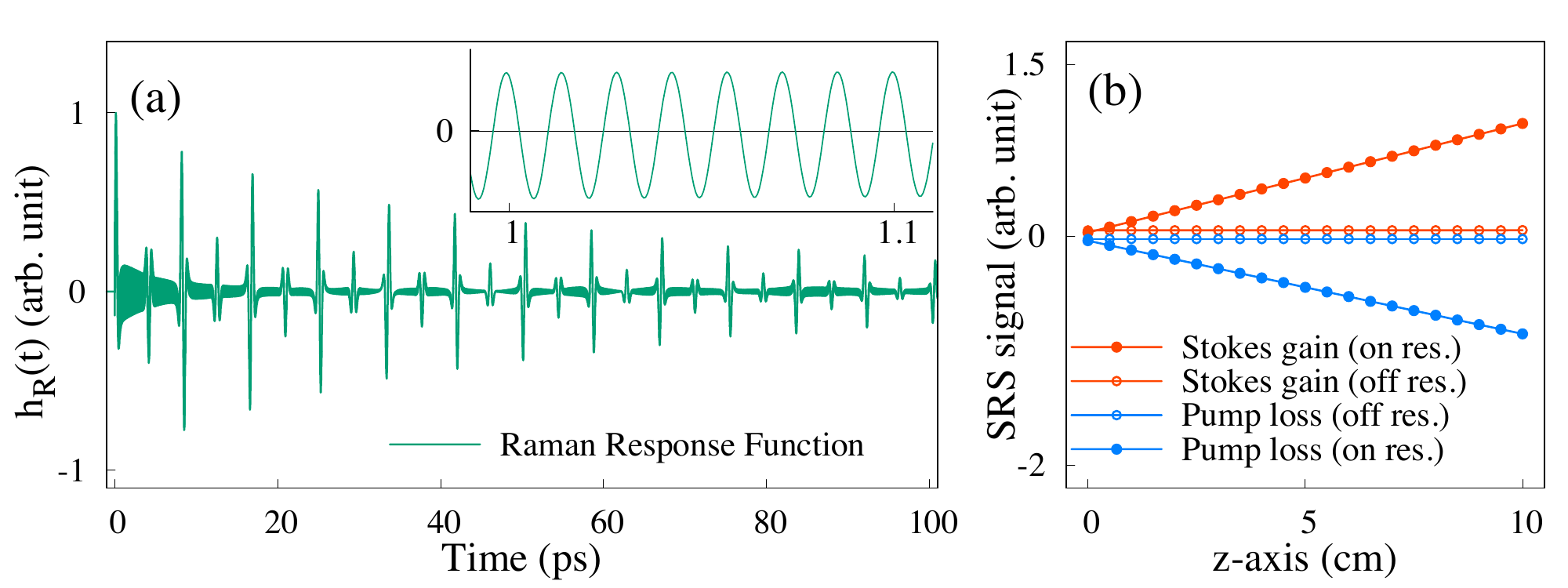}
    
    \caption{(a) Raman response function of molecular nitrogen including the rovibrational Q-branch transitions and the rotational transitions. The insert shows the fast oscillation from 1 ps to 1.1 ps. (b) SRS signal (SRG and SRL) versus the propagating distance along the z-axis. The points with solid circles represent the resonant case and the points with open circles for the non-resonant case.}
    \label{fig:RamanRes}
    \end{figure*}

      Fig. \ref{fig:RamanRes} (a) shows the Raman response function of molecular nitrogen over time. 
    The high-frequency oscillations shown in the insert figure are from the rovibrational Q-branch transitions around 2330 $cm^{-1}$. 
    The full revival with a period of 8.4 ps originates the rotational transitions. 
    The Raman response function presented in Fig. \ref{fig:RamanRes} (a) is perfectly consistent with recent experimental measurements with very high sensitivity \cite{2015Peng}. The detailed parameters of the Raman response function for molecular nitrogen and oxygen could be found in \cite{2007Zheltikov,Long2002,2000Bendtsen}. 
    Fig. \ref{fig:RamanRes} (b) shows the variation of laser intensity for Stokes burst (red) and pump burst (blue) as a function of the propagating distance in space. In the simulation, the inter-pulse temporal separation ($\Delta \tau$) is 1 ps, and the number of pulses ($N$) is 100.
    As illustrated in Fig. \ref{fig:1}, by changing the offset phase, one can tune the SRS between on-resonance (solid circles) and off-resonance (open circles).
    For the resonant offset phases, the pulse energy can sufficiently flow from the pump burst to the Stokes burst, resulting in their loss and gain respectively.
    The linear growth of intensity over propagation distance is consistent with the formula deduced in \cite{2018Polli} in the weak signal limit.
    % Data were slightly shifted from 0 position in y-axis for better visualization.

    \subsection{Raman shift scanning}
    
    % \begin{figure*}[htbp]
    \begin{figure*}[ht!]
    \centering
    
    \includegraphics[width=0.85\textwidth]{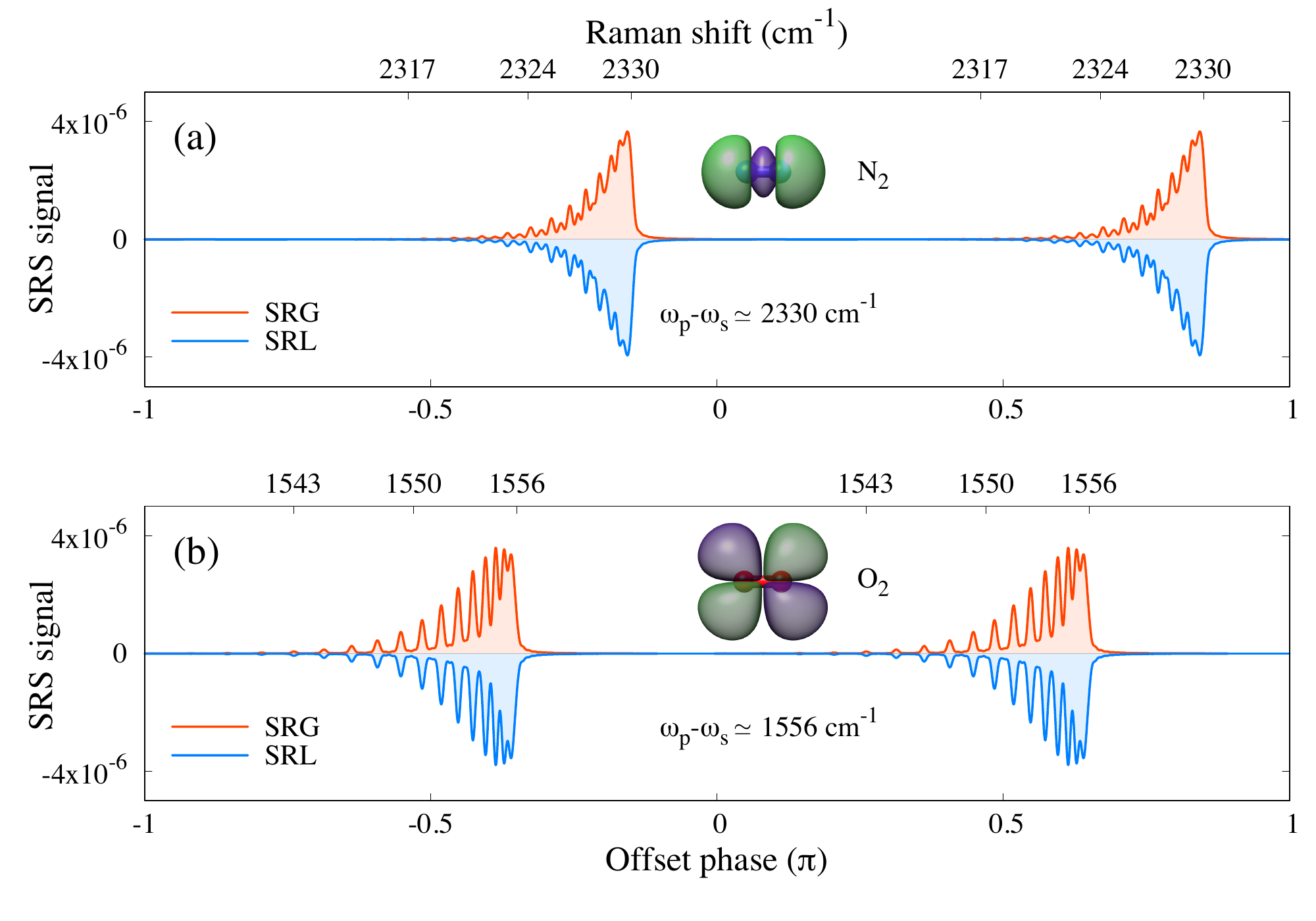}
    
    \caption{Raman shift spectrum of burst-driven SRG (red curve) and SRL (blue curve) around 2330 $cm^{-1}$ in $N_2$ (a) and around 1556 $cm^{-1}$ in $O_2$ (b) by changing the offset phase from $-\pi$ to $\pi$. The upper axis is the Raman shift energy in $cm^{-1}$ corresponding to the offset phase. In those simulations, $N=100$, $\Delta \tau=1ps$, $I_s/I_p = 0.8$.}
    \label{fig:repeat}
    \end{figure*}

    Fig. \ref{fig:repeat} shows the calculated Raman shift spectrum of Q-branch ($\Delta J=0$) associated with the vibrational transition $v=0 \rightarrow v=1$ of molecular nitrogen (a) and molecular oxygen (b) in their ground states $X ^1\Sigma_g^+$ and $X ^3\Sigma_g^-$ respectively. 
    With the present spectral resolution, $(N\cdot\Delta\tau)^{-1}= 0.33 cm^{-1}$, the $J$ state ratio of both molecules are clearly shown in the Raman shift spectra. $N_2$ has a nuclear spin quantum number of 1, so its even-$J$ to odd-$J$ ratio is 2:1. The nuclear spin quantum number for $O_2$ is 0, and its even-$J$ is forbidden.
    The simulations shown in Fig. \ref{fig:repeat} are nicely agreed with previous experimental results \cite{Long2002,2000Bendtsen}. 
    The expected periodic repetition of every $\pi$ is also revealed in Fig. \ref{fig:repeat}, which is a unique feature of BSRS and indicates an easy phase-to-frequency calibration procedure in practice.
    The signal yield ratio between SRG and SRL can also help the calibration procedure and could be estimated by $\frac{SRG}{SRL} = \frac{I_p \omega_s}{I_s \omega_p}$, which is $\simeq1$ under current considerations.

\section{Discussion} \label{sec:ResDis}

    % ====================== Para 4: Resolution vs N ================
    \subsection{Spectral resolution}
    
    \begin{figure}[htbp]
    \centering
    \includegraphics[width=0.95\textwidth]{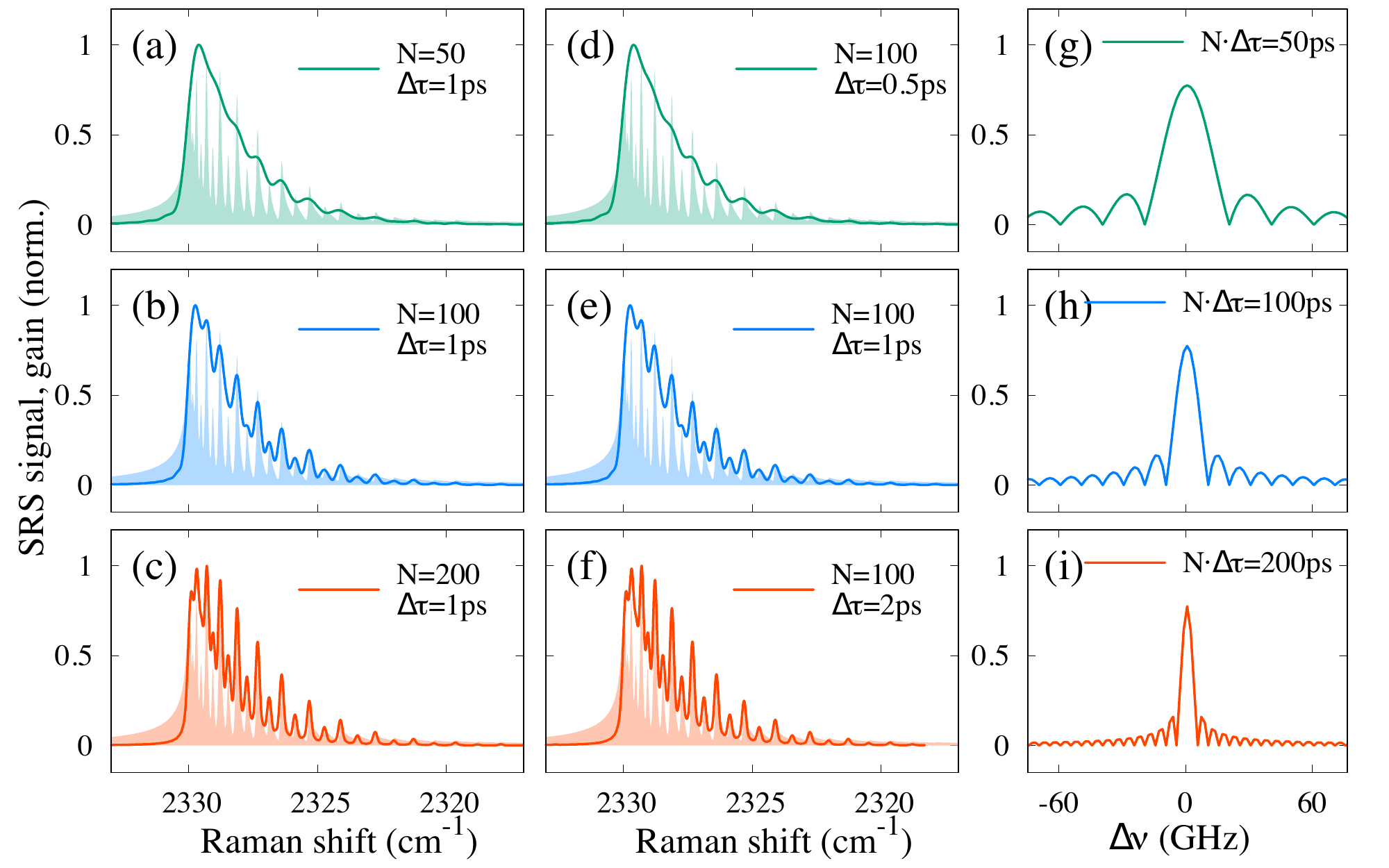}
    \caption{Normalized Raman shift spectrum dependence on the number and inter-pulse temporal separation of pulses in burst: (a, d) $N \cdot \Delta \tau$ = 50 ps, (b, e) $N \cdot \Delta \tau$ = 100 ps and (c, f) $N \cdot \Delta \tau$ = 200 ps. (a-c) changing the number of pulses, keeping the inter-pulse temporal separation unchanged. (d-f) changing the inter-pulse temporal separation and keeping the number of pulses unchanged. The filled curve in (a-f) is the Fourier Transformation of the Raman Response function of $N_2$. (g-i) Width of individual peaks in the comb for different values of $N \cdot \Delta \tau$: (g) $N \cdot \Delta \tau$ = 50 ps, (h) $N \cdot \Delta \tau$ = 100 ps, and (i) $N \cdot \Delta \tau$ = 200 ps. The filled areas in (a-f) present the Raman spectrum from the direct Fourier transform of the Raman response function.}
    \label{fig:resolution}
    \end{figure}
    
    The high spectral resolution is desired to separate the species if their Raman shift spectra distributions are close to or overlap with each other. 
    For BSRS, the spectral resolution is given by the reciprocal of the product of the number of pulses and the inter-pulse temporal separation, $(N \cdot \Delta\tau)^{-1}$, which means increasing either $N$ or $\Delta \tau$ can improve the spectral resolution.
    The results shown in Fig. \ref{fig:resolution} (a-f) confirm the spectral resolution formula. It can be seen that as $N \cdot \Delta\tau$ increases from 50 ps to 200 ps, the spectral peaks become narrower, and the distinction between even J and odd J becomes more pronounced.
    The corresponding spectral resolution of (a, d) is 0.66 $cm^{-1}$, (b, e) is 0.33 $cm^{-1}$, and (c, f) is 0.17 $cm^{-1}$.
    Fig. \ref{fig:resolution} (g-i) shows the spectral width of the individual peak in comb getting narrower and narrower when $N \cdot \Delta\tau$ increases from 50 ps to 200 ps, as predicted in the previous part. It indicates that the larger product of $N$ and $\Delta \tau$ corresponds to a narrower spectral bandwidth leading to a higher spectral resolution.

    A single-pulse driven field with a full-width half max (fwhm) duration of 200 ps can also achieve such high spectral resolution, however its peak intensity is limited by the ps laser technology.
    On the one hand, the burst has a higher laser intensity than the single pulse field when the total pulse energy is the same. For example, the laser intensity of a burst with $N=200$ and fwhm=100 fs is ten times higher than that of a signal pulse with fwhm=200 ps when they have the same total laser pulse energy.
    On the other hand, the chirped pulse amplification technique allows more energy to be put into fs bursts.
    As SRS is a non-linear process, its signal yield benefits from a higher laser intensity. SRS driven by a higher intensity field has higher detection sensitivity, which can detect molecules of interest with lower counts in the sample and requires a shorter sample thickness to obtain the same signal yield, and more importantly, avoid heating effects that may damage the sample.

    % ====================== Para 6: signal enhancement (new results) ================
\subsection{Resonant dynamics}

 \begin{figure}[htbp]
    \centering
    \includegraphics[width=1\textwidth]{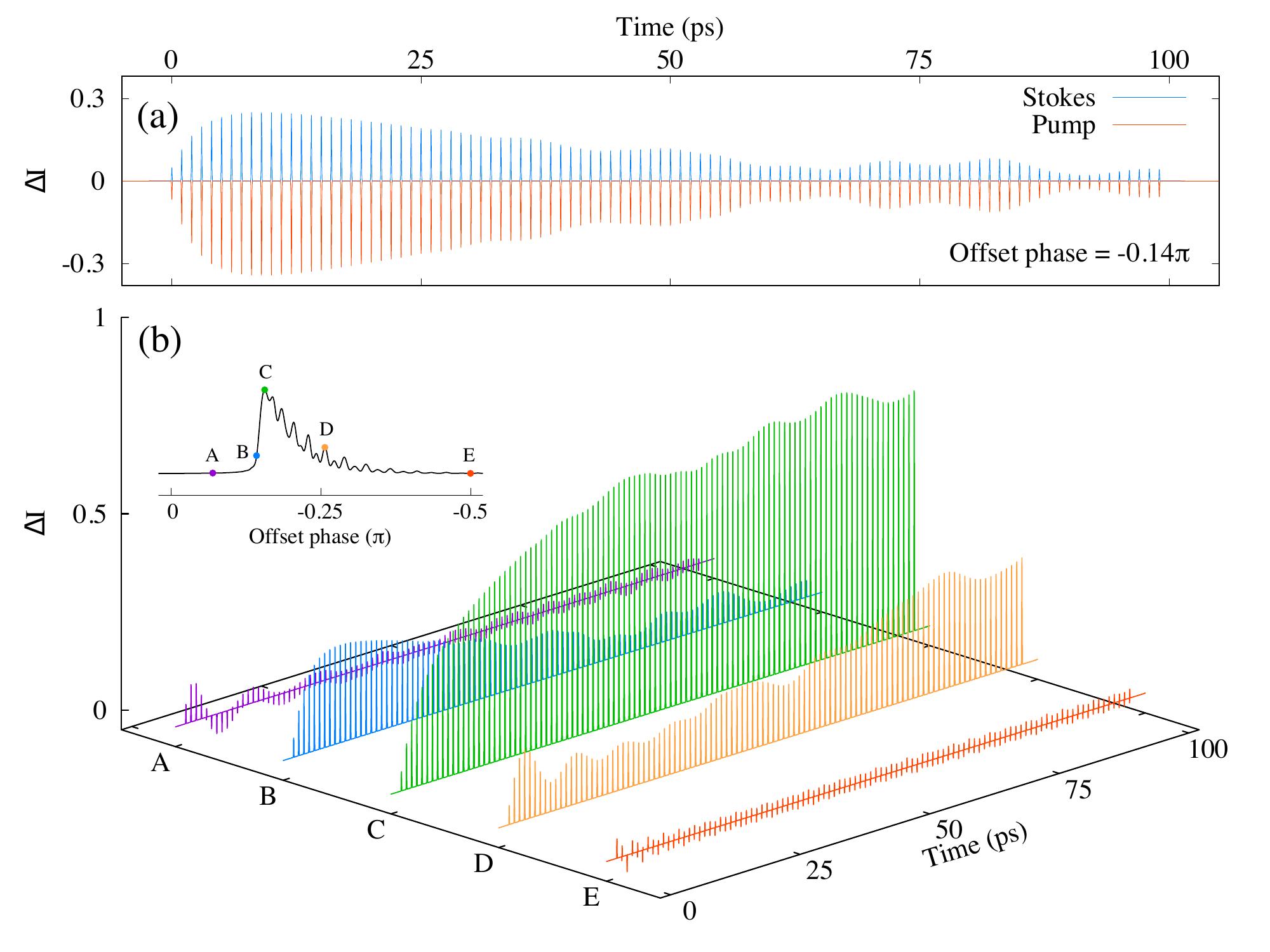}
    \caption{Intensity difference between the driven burst before and after the Stimulated Raman process, $\Delta I_{s,p}(t) = |A_{s,p}(z,t)|^2-|A_{s,p}(z=0,t)|^2$. (a) $\Delta I_{s}(t)$ and $\Delta I_{p}(t)$ as a function of time when the offset phase is $-0.14 \pi$. (b) $\Delta I_{s}(t)$ for different offset phases. The offset phases corresponding to the points are shown in the insert figure, their values are A = $-0.07 \pi$, B = $-0.14 \pi$, C = $-0.16 \pi$, D = $-0.26 \pi$, E = $-0.5 \pi$. }
    \label{fig:PhaseScan}
    \end{figure}

This part is to intuitively understand the onset and end of resonance dynamics of BSRS. Fig. \ref{fig:PhaseScan} shows the laser intensity difference of driven burst ($N=100, \Delta \tau=1ps$) before and after the Stimulated Raman process, $\Delta I_{s,p}(t) = |A_{s,p}(z,t)|^2-|A_{s,p}(z=0,t)|^2$. 
$\Delta I(t)$ for the Stoke burst (red curve) and the pump burst (blue curve) are depicted in Fig. \ref{fig:PhaseScan} (a), in which their coupling could be observed obviously.
Due to the conservation of photon numbers, the equation $\frac{\Delta I_s(t)}{\omega_s} + \frac{\Delta I_p(t)}{\omega_p}  =0$ holds all the time. Thus, $\Delta I_s(t)/\Delta I_p(t)$ shown in Fig. \ref{fig:PhaseScan} (a) is around 0.8 for the whole time domain.

Fig. \ref{fig:PhaseScan} (b) explains how the offset phase affects the net gain by showing $\Delta I_s(t)$ for different values of the offset phase corresponding to Point A-E shown in the insert figure. 
It shows that the gain and loss of driven laser fields happen in every pulse, which is reasonable since each pair of fs-pulses in Stoke and pump bursts can induce the SRS process.
However, by integrating $\Delta I(t)$ over time, it is found the net gain and loss only exist for the resonant offset phases, like Point B, C and D. For the non-resonant offset phases, like Point A and E, the negative and positive parts completely cancel out, leading to zero Raman signal. 
The range and value of the net gain are mainly determined by the value of the offset phase. For example, from Point A to Point C, the range of net gain grows from a few ps to 200 ps, and a more than tenfold increase in value.
Point C has the strongest resonance, every pulse in the burst constructively contributes to the net gain at a high level of value. At this point, the Raman molecular vibration induced by each pair of pumps and Stokes pulse can constructively superimpose and thus give a strong net gain all over the whole burst.

\subsection{Signal yield}

  \begin{figure}[htbp]
    \centering
    \includegraphics[width=0.9\textwidth]{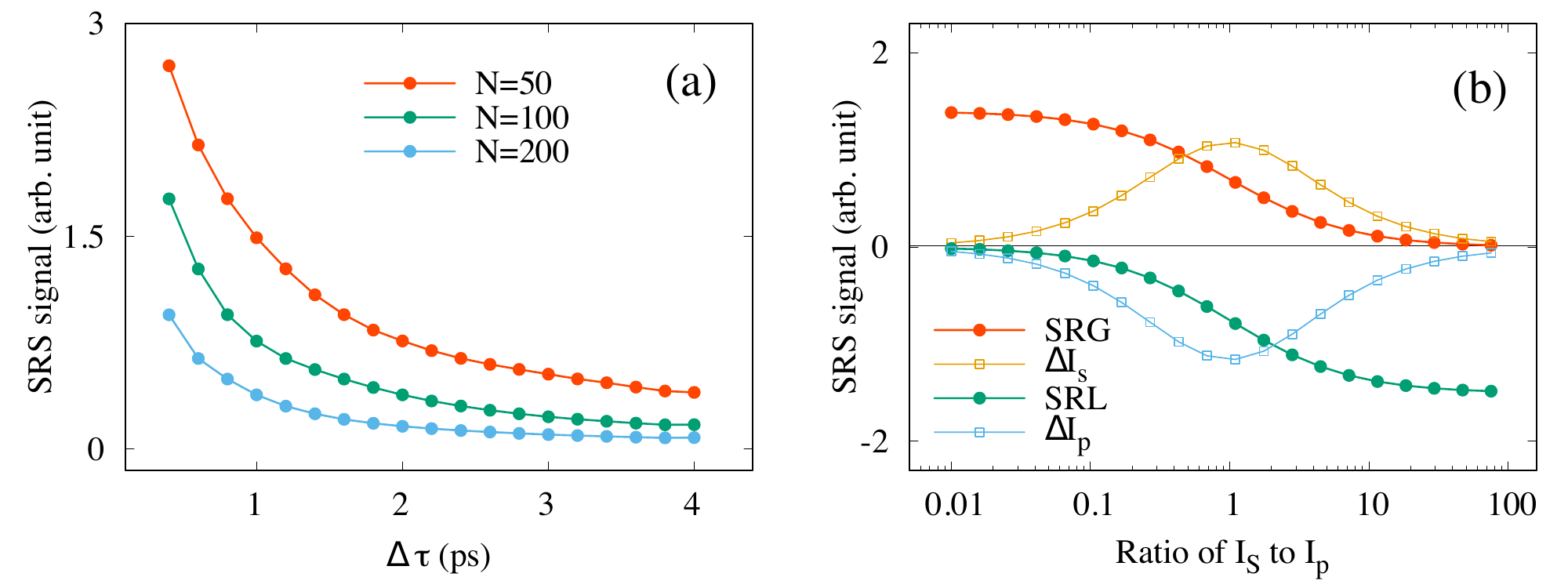}
    \caption{Dependence of Raman signal yield on driving laser burst parameters: (a) number of pulses in burst and the inter-pulse temporal separation; (b) laser intensity ratio of Stokes burst to pump burst. The total pulse energy is fixed for those results.}
    \label{fig:tau}
    \end{figure}

Fig. \ref{fig:tau} presents the effect of the driven laser parameters, including the number of pulses $N$, the inter-pulse temporal separation $\Delta \tau$, and the intensity ratio $I_s/I_p$, on the stimulated Raman signal. 
The total laser pulse energies are the same for all simulations in Fig. \ref{fig:tau}. 
In Fig. \ref{fig:tau} (a), the blue line-points ($N=200$) are below the red line-points ($N=50$) which indicates the signal strength decreases as $N$ increases. The reason is that a larger $N$ leads to a smaller energy per pulse in the burst and a lower laser intensity resulting in a decrease in the signal yield. 
We also see increasing $\Delta\tau$ leads to a decrease in yield for every value of $N$. The decrease of fast oscillations around 0$\sim$10 ps of the Raman response function in Fig. \ref{fig:RamanRes} (a) give an explanation: as $\Delta\tau$ increasing, the pulse-to-pulse superposition becomes less effective which reduces the efficiency of SRS.
A larger value of $N\cdot\Delta\tau$ means higher spectral resolution and lower SRS signal yield, thus Fig. \ref{fig:tau} (a) tells us that when the total laser pulse energy is fixed, we can only get a high spectral resolution with relatively lower SRS signal yield, or the high SRS signal yield with relatively lower spectral resolution.
% Another feature of Fig. \ref{fig:RamanRes} (a) is that for N=200, the signal yield does not change a lot when $\Delta\tau$ is larger than 1 ps, while for N=100, $\Delta\tau$ must be larger than 2 ps. \xinhua{This part is not clear.} 
% The reason also lies in the Raman response function. The Raman response function nearly decreases to zero within 200 ps, the changing of signal yield is not obvious when the total burst width is larger than 200 ps.
$\Delta\tau$ can affect the Raman scanning range which equals to $\frac{33.3 cm^{-1}}{\Delta\tau[ps]}$, so $\Delta\tau$=4ps corresponds to a scanning range of 7 $cm^{-1}$. Therefore, we only calculate $\Delta\tau$ scanning up to 4ps to cover the rovibrational spectrum of $N_2$ of interest.

The effect of laser intensity ratio to the signal yield is depicted in Fig. \ref{fig:tau} (b), in which $\Delta \tau$ = 1 ps and $N$ = 100. We observe that SRG = $\Delta I_s/I_s$ is higher when $I_p \gg I_s$, while SRL =$\Delta I_s/I_s$ prefers $I_p\ll I_s$, which is consisted with \cite{2018Polli}. 
The open square points show the absolute changes of $\Delta I_s$ and $\Delta I_p$. 
The peak position for them locate at $I_s = I_p$.
A previous research  \cite{2015Moester} showed that to achieve a high signal-to-noise ratio, the laser intensity ratio should be set as $I_p/I_s=2$ for SRG and $I_s/I_p=2$ for SRL.

 \begin{figure}[htbp]
    \centering
    \includegraphics[width=0.9\textwidth]{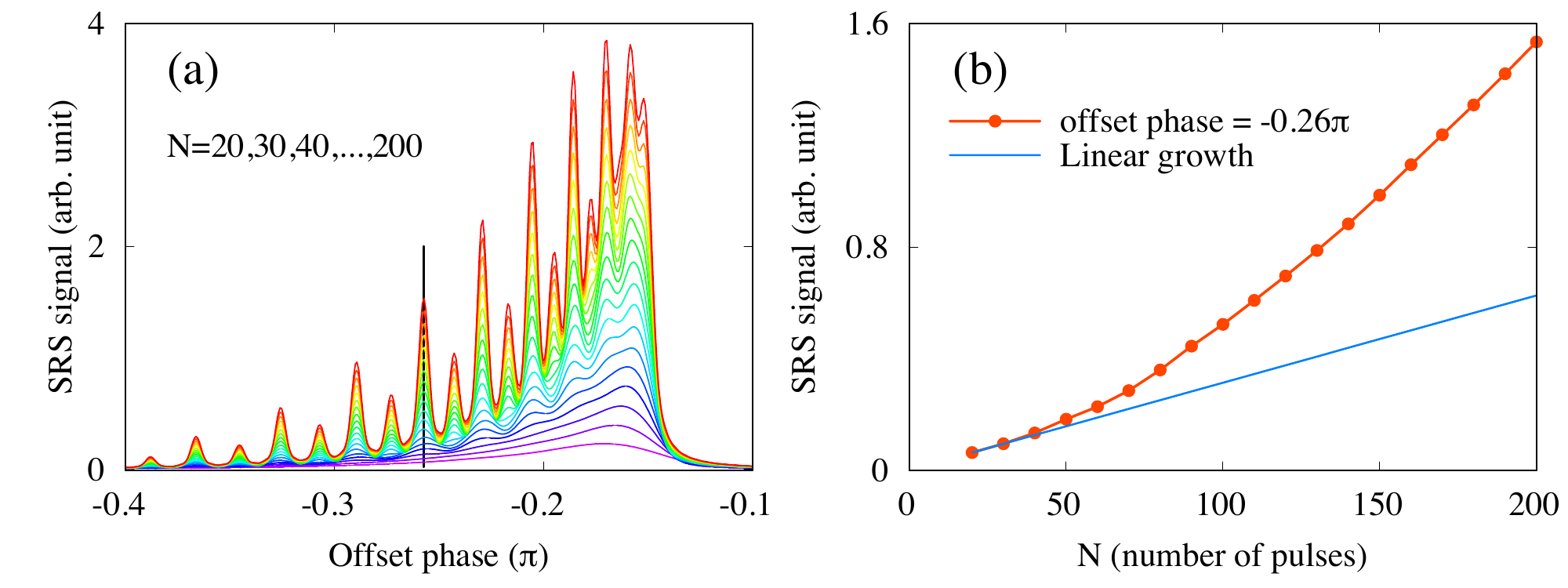}
    \caption{Raman signal yield versus number of pulses ($N$) for the fixed laser intensity. (a) the corresponding N from bottom to top is 20, 30,..., 200. (b) is the data of the offset phase equals to -0.26 $\pi$, indicated by the vertical black line in (a). $\Delta \tau$ is 1 ps in those simulations.}
    \label{fig:growth}
    \end{figure}

In addition to the signal yields for a fixed total laser pulse energy of bursts, as shown in Fig. \ref{fig:tau}, we have also investigated the signal yield as a function $N$ when the laser intensity is fixed, which means the total laser pulse energy of bursts increases as $N$ increasing, to reveal the coherent growth laws. 
Fig. \ref{fig:growth} (a) shows the Raman shift spectra for $N$ increasing from 20 to 200. It can be seen that the resolution increases as $N$ increases. 
For $N$=20, the Raman shift spectrum (bottom purple curve) has a broad peak that can not separate the resonance or non-resonance narrow peaks. 
For $N$=200, the Raman shift spectrum (top red curve) clearly shows the rovibrational structures of the nitrogen, and the odd-$J$ and even-$J$ can be distinguished easily. 

Another feature is that as $N$ increases, the signal yield becomes larger, especially at the resonant frequency of the nitrogen.
The red dotted line in Fig. \ref{fig:growth} (b) shows the signal yield growth at offset phase = -0.26 $\pi$ indicated by the vertical black line in Fig. \ref{fig:growth} (a).
The reason for faster than linear growth (blue line) is the offset phase at -0.26 $\pi$ corresponds to the resonance frequency of the $N_2$, thus the signal yield grows coherently when increasing $N$ due to constructive interference of the signals from each pulse pairs in the bursts.
In Eq. \ref{equ:ramanfunc}, the sum part $\sum_{m=0}^N \text{exp}(im\alpha)$ is equal to $\frac{sin(N\alpha/2)}{sin(\alpha/2)}e^{-i(N-1)\alpha/2}$, where $\alpha$ is set as $\Omega\cdot\Delta\tau-2\phi$. 
In the experiment, the measured signal is the laser intensity, so we need to investigate the signal growing properties with $N$ by the square of the sum part, which is $|\sum_{m=0}^N e^{im\alpha}|^2 = \frac{sin^2(N\alpha/2)}{sin^2(\alpha/2)}$. 
And when $\alpha\rightarrow0$, we have the pure quadratic growth with $N$ as $\lim _{\alpha\rightarrow0}\frac{sin^2(N\alpha/2)}{sin^2(\alpha/2)} = N^2 $.
Such a condition can be achieved when the Raman response function consists of only one frequency without any relaxation time, which corresponds to a Delta function in the frequency domain.
However, the real Raman response function contains different frequencies with a certain width, the Raman signal includes also the contribution from the off-resonant frequency components, which grows lower than the quadratic.
Therefore, the signal growth over $N$ shown in Fig. \ref{fig:growth} (b) exhibits rather a mixture of quadratic and linear growth.
The nonlinear growth of the signal yield over $N$ indicates that a higher signal-to-noise ratio can be achieved with the BSRS as compared to those methods with single pulses.

\subsection{Pixel dwell time}

    \begin{figure}[htbp]
    \centering
    \includegraphics[width=0.6\textwidth]{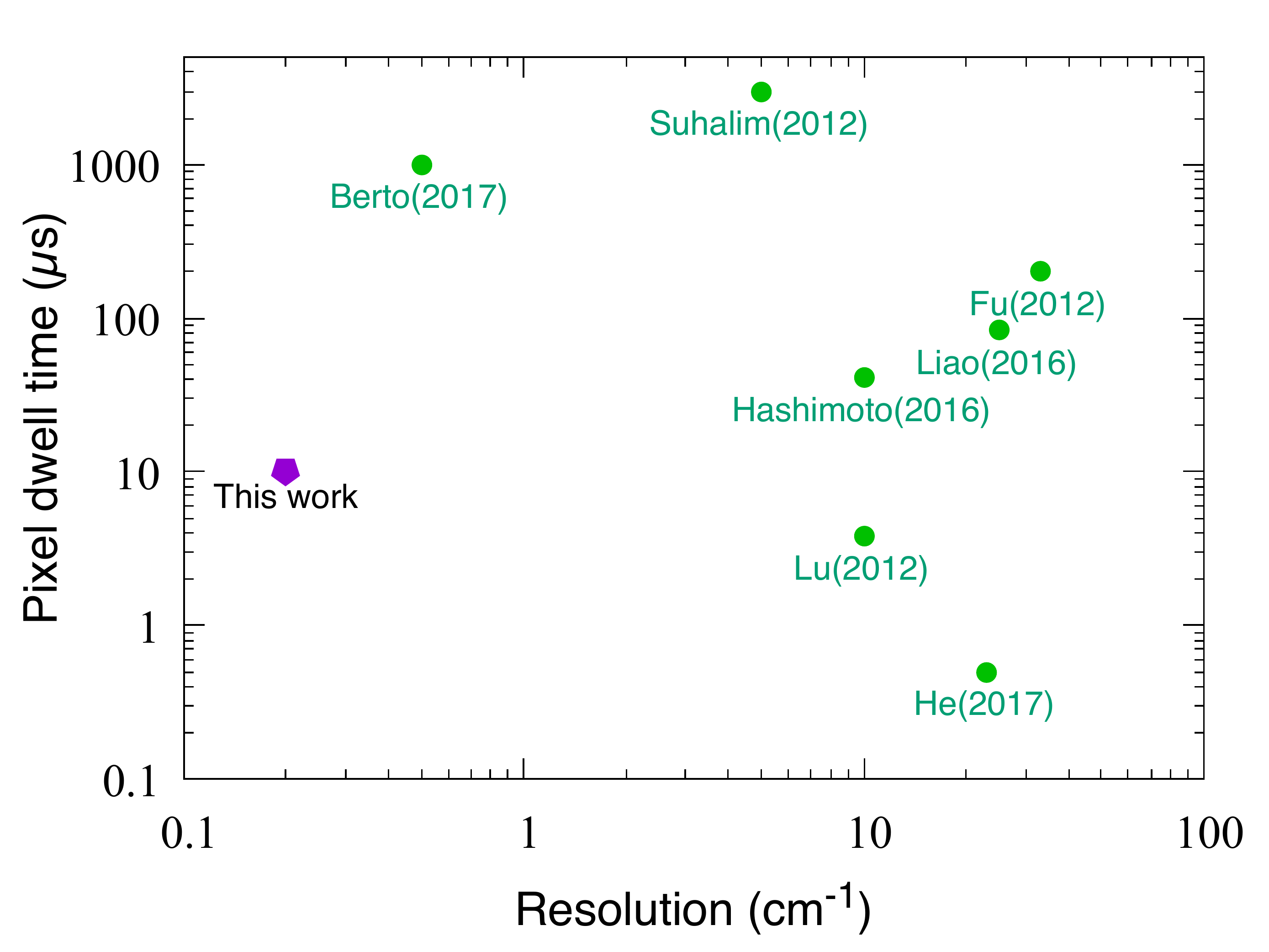}
    \caption{ Comparison of the stimulated Raman spectroscopy methods on spectral resolution and pixel dwell time. Suhalim (2012) \cite{2012Suhalim}, Berto (2017) \cite{2017Berto}, Fu (2012) \cite{2012Fu1}, Liao (2016) \cite{2016Liao}, Hashimoto (2016) \cite{2016Hashimoto}, Lu (2012) \cite{2012Lu}, He (2017) \cite{2017He}.  }
    \label{fig:overview}
    \end{figure}
    
Pixel dwell time is the minimum time the driven laser should interact with the sample for a single spectral pixel or spatial pixel. It is directly relative to the data acquisition speed: shorter pixel dwell time means higher data acquisition speed. 
High data acquisition speed is particularly needed in the studies of dynamic events in a living system \cite{2021ReviewLiu}. 
In our method, the offset phase is controlled by an AOM working at the frequency of 300$\sim$400 MHz. Considering the MO working frequency is around 100 MHz, AOM actually could achieve the intensity and phase controlling of each pulse one by one, and thus can achieve the controlling of input phase burst by burst. As an example, if RA works at the frequency of 100 kHz, the pixel dwell time of this approach can reach down to 10 $\mu s$.
Fig. \ref{fig:overview} shows the overview of the broadband coherent Raman spectroscopy methods in the spectral resolution domain and pixel dwell time domain \cite{2018Polli}. Combining with the advantages of fs pulses and ps pulses, this work achieves the product of spectral resolution and pixel dwell time as low as \(2 ~ \mu s\cdot cm^{-1}\). 
This is to the best of our knowledge the first method to achieve such performance.

To realize this unique method, several possible technical challenges in experiment need to be considered. 
Currently, a burst with 6 pulses has already been generated in our lab \cite{2020Vinzenz}, while to achieve a high spectral resolution $N$ should be increased to 100. Since the first pulse and the last pulse in the burst have different round-trip numbers in the optical amplifier, the dispersion effects may need consideration.
Secondly, the choice of $f_R$. $f_R$ refers to the magnitude fraction between the nonresonant background signal to the stimulated Raman signal.
% We should pay attention to choosing the value of it.
In our simulation, $f_R$ was set as 0.95 based on the fact that the resonant signal yield is a mixture of quadratic growth and linear growth, which is much higher than the frequency-irrelative non-resonant signals. 
The exact value of $f_R$ is related to the signal-to-noise ratio and can be determined by fitting the measured experimental results.
By overcoming those technical issues, BSRS can become a powerful and robust tool with high spectral resolution, high speed, high sensitivity and high accuracy. It will greatly benefit applications ranging from biological imaging \cite{2015ReviewXie}, environmental gas sensing \cite{2011Hemmer}, materials characterization, and other fields based on SRS.

In conclusion, we demonstrate numerically a novel approach to achieve the Stimulated Raman spectrum with high spectral resolution, high-speed acquisition, and high sensitivity of detection by using fs-pulse bursts. 
Our approach promises high sensitivity and efficiency simultaneously since it requires neither wavelength detuning and moving delay stages, nor precise dispersion management. 
We additionally analyze the underlying dynamics of SRS driven by fs bursts for the resonance and non-resonance cases, the signal yield dependence on the parameters of driven laser fields.
We expect this approach to significantly expand the application of SRS process in various areas.

\section{Materials and methods}

    The critical point to validate this technique is to prove that the broadband SRS signal, arising from the interaction with each respective pair of fs signal and idler pulses in the time-coincident bursts, will retain the expected spectral resolution given by $(\Delta \tau\cdot N)^{-1}$.
    Using molecular nitrogen as a model system, we present systematic numerical simulations, based on the coupled nonlinear Schrödinger equations \cite{2013BookAgrawal} shown in Eq. (\ref{equ:NLSE1}). A detailed procedure for the derivation of the coupled nonlinear Schrödinger equations can be found in the literature \cite{1996Headley,1995Headley}.
    \begin{equation} \label{equ:NLSE1}
        \begin{split}
        \frac{\partial A_p}{\partial z} & =i\gamma_p(1-f_R)(|A_p|^2A_p+2|A_s|^2A_p) +R_p(z,t) \\
        \frac{\partial A_s}{\partial z} & =i\gamma_s(1-f_R)(|A_s|^2A_s+2|A_p|^2A_s) +R_s(z,t)
        \end{split}
    \end{equation}
    In the equations, $A_p$ and $A_s$ are the slowly varying envelopes associated with the pump and Stokes pulse bursts, $\gamma_{p,s}$ is the nonlinear parameters, $f_R$ represents the fractional contribution of the resonant Raman response to nonlinear polarization \cite{1992Stolen,2018Polli,2013BookAgrawal,1997Ripoche}. 
    % In our simulation, $f_R$ was set around 0.9, since the self-phase modulation, the first term on the right side of Eq. (\ref{equ:NLSE1}), is relatively weak for species in gas states for its small the second-order nonlinear refractive index, $n_2$ \cite{2011Geints}. 
    $R_{p,s}(z,t)$ shown in Eq. (\ref{equ:ramanfunc}) is the Raman contribution term, which is a function of the frequency detuning ($\Omega$) and Raman response function [$h_R(t)$]. 
    The Raman response function of molecular nitrogen \cite{2007Zheltikov,Long2002} is depicted in Fig. \ref{fig:RamanRes} (a).

    % f_R description in 2013BookAgrawal {Agrawal book Eq. (2.3.39) page 41}
    % nonlinear description in 2018Polli
    
    \begin{equation} \label{equ:ramanfunc}
        \begin{split}
          R_{p,s}(z,t)=i\gamma_{p,s}f_R  A_{p,s} \int_{-\infty}^t dt' h_R(t-t')\left( |A_{p,s}(z,t')|^2+ |A_{s,p}(z,t')|^2\right) +\\
          i\gamma_{p,s}f_R  A_{s,p} \int_{-\infty}^t dt' h_R(t-t')A_{p,s}(z,t')A_{s,p}^*(z,t')\exp(\pm i\Omega(t-t'))\sum_{m=0}^N\exp(im(\Omega\cdot\Delta\tau-2\phi))
        \end{split}
    \end{equation}

    The electric field of the pump and Stokes pulse bursts, $E_{p,s} = A_{p,s} \cos(\omega_{p,s}t+\Phi_{p,s}) $, are given by Eq. (\ref{equ:bursts}), with the laser intensity ($I_{p,s}$) around $10^{11\sim12}$ $W/cm^2$, the pulse duration ($v_{p,s}$) of individual pulses in burst equals to 100 $fs$. 
    % $\Omega=\omega_p-\omega_s $ is the Stokes shift.
    $\omega_p$ and $\omega_s$ are the respective central carrier frequency of the broadband individual pulse in the pump burst and Stokes burst, which is approximately detuned to the Raman shift frequency, $\Omega_R \approx \omega_p-\omega_s $. The frequency detuning is $\omega_p-\omega_s $, denoted by $\Omega$.
    $\lambda_p$ = 919 $nm$ and $\lambda_s$ = 1170 $nm$ were chosen such that their angular frequency difference locates in the range of the Raman shift (2330 $cm^{-1}$) of molecular nitrogen and the wavelength corresponding to their angular frequency sum is about 1030 $nm$ - the central wavelength of Yb-doped lasers \cite{2009Pugzlys}.
    Additional to molecular nitrogen, we also performed simulations with molecular oxygen using the same model. For molecular oxygen, the Raman shift is around 1556 $cm^{-1}$, and the wavelengths were set as $\lambda_p$ = 965 $nm$ and $\lambda_s$ = 1135 $nm$. 
    Due to constructive conditions between pulses in the burst, the offset phase ($\phi$) and the inter-pulse temporal separation ($\Delta\tau$) are relative to the frequency detuning naturally.
    The Stimulated Raman Gain (SRG) and Stimulated Raman Loss (SRL) are given by $SRG=g_s=\frac{\Delta I_s}{I_s}$ and $SRL=l_p=\frac{\Delta I_p}{I_p}$, respectively.

    % $\Omega=(\omega_p-\frac{\phi}{2\pi \Delta \tau})-(\omega_s+\frac{\phi}{2\pi \Delta \tau}) \simeq 2323.7 cm^{-1}-\frac{\phi}{\pi}\text{33.3 }cm^{-1} $
    
    % \begin{equation} \label{equ:bursts}
    %     \begin{split}
    %       E_p(t)&=\sum_{n=0}^{N}{\sqrt{I_p}\exp\left[-\frac{(t-n\cdot\Delta \tau)^2}{(\tau_p/2)^2}\right]\cos(\omega_p\cdot t-n\cdot \phi)}\\
    %       E_s(t)&=\sum_{n=0}^{N}{\sqrt{I_s}\exp\left[-\frac{(t-n\cdot\Delta \tau)^2}{(\tau_s/2)^2}\right]\cos(\omega_s\cdot t+n\cdot \phi)}
    %     \end{split}
    % \end{equation}
    
    \begin{equation} \label{equ:bursts}
        \begin{split}
          E_p(t)&=\sum_{m=0}^{N}{\sqrt{I_p}\exp\left[-\frac{(t-m\Delta \tau)^2}{(v_p/2)^2}\ln2 \right]\cos(\omega_p (t-m\Delta \tau)-m\phi)}\\
          E_s(t)&=\sum_{m=0}^{N}{\sqrt{I_s}\exp\left[-\frac{(t-m\Delta \tau)^2}{(v_s/2)^2}\ln2 \right]\cos(\omega_s (t-m\Delta \tau)+m\phi)}
        \end{split}
    \end{equation}
We know from the working principle that the frequency detuning can be realized through simply changing the offset phase ($\phi$). 
    Based on the interference conditions of pulses in burst, the frequency detuning can be estimated by the formula of $\Omega_{detuning}=\left(\frac{ M_p\cdot 2\pi}{\Delta \tau}+\frac{\phi}{ \Delta \tau}\right)-\left(\frac{M_s \cdot 2\pi }{\Delta \tau}-\frac{\phi}{\Delta \tau}\right)$, where $M_{p,s}$ is the nearest integer of $\frac{\omega_{p,s}\cdot \Delta\tau}{2\pi}$.
    This formula indicates that the scanning range is determined by both the central frequency difference ($\omega_p-\omega_s$) and the inter-pulse temporal separation ($\Delta \tau$). 
    $\omega_p-\omega_s$ can be used to determine the general scanning range. 
    % For instance, the central wavelengths of the pump burst and the Stoke burst were set as 919 $nm$ and 1170 $nm$ for $N_2$, and $\lambda_p$ = 965 $nm$ and $\lambda_s$ = 1135 $nm$ for $O_2$.
    This is quite useful for detecting species with the Raman shift spectrum in different spectral ranges.
    $\Delta \tau$ defines fine scanning range adjustments. For example, $\Delta \tau$=1 ps corresponds to 33 $cm^{-1}$. This is sufficient to measure the rovibrational energy levels of $N_2$ and $O_2$ in the range of about 10 $cm^{-1}$.
    % It is rather efficient and does not require wavelength detuning and precise dispersion management of the laser pulses, and moving delay stages, or forming spectral windows. 
    % Its efficiency and the pixel dwell time will be discussed later.

% \backmatter

% \bmhead{Supplementary information}

\bmhead{Acknowledgements}
{This work was supported by Austrian Science Fund (FWF) under ZK 9100-N and I 4566.}

\bmhead{Author details}
{$^1$ Photonics Institute, Technische Universität Wien, Gußhausstraße 27-29, A-1040 Vienna, Austria
$^2$ SwissFEL, Paul Scherrer Institute, Villigen PSI 5232, Switzerland
$^3$ Department of Physics and Astronomy, Texas A$\&$M University, College Station, Texas 77843, USA}

\bmhead{Author contributions} 
{A.B. initiated this project.
H.H. performed the simulation with the help from X.X., A.Z., and A.B..
H.H. X.X. and A.B. analyze the results, and all the authors contributed to the discussion on the results.
H.H. and X.X. write the manuscript with assistance from all the authors.}

\bmhead{Data availability}
{All the data and methods needed to evaluate the conclusions of this work are present in the main text. Additional data can be requested from the corresponding author.}

\bmhead{Conflict of interest}
{The authors declare no conflicts of interest.}

%%===========================================================================================%%
%% If you are submitting to one of the Nature Portfolio journals, using the eJP submission   %%
%% system, please include the references within the manuscript file itself. You may do this  %%
%% by copying the reference list from your .bbl file, paste it into the main manuscript .tex %%
%% file, and delete the associated \verb+\bibliography+ commands.                            %%
%%===========================================================================================%%

% \bibliography{sn-bibliography}% common bib file
%% if required, the content of .bbl file can be included here once bbl is generated
%%\input sn-article.bbl

\bibliography{refs}

\end{document}